\documentclass{emulateapj}\usepackage{psfig,graphicx,natbib, float}
\shorttitle{Granulation as an astrophysical noise source}
\shortauthors{Cegla et al.}

\begin{document}
\title{Stellar Surface Magneto-Convection as a Source of Astrophysical Noise\\ I. Multi-component Parameterisation of  Absorption Line Profiles}
\author{H.~M. Cegla\altaffilmark{1,2}, S. Shelyag\altaffilmark{1}, C.~A. Watson\altaffilmark{1}, M.\ Mathioudakis\altaffilmark{1}}
\altaffiltext{1}{Astrophysics Research Centre, School of Mathematics \& Physics, Queen's University, University Road, Belfast BT7 1NN, UK; {\tt hcegla01@qub.ac.uk}}
\altaffiltext{2}{Department of Physics \& Astronomy, Vanderbilt University, Nashville, Tennessee 37235, USA}

\begin{abstract}
We outline our techniques to characterise photospheric granulation as an astrophysical noise source. A four component parameterisation of granulation is developed that can be used to reconstruct stellar line asymmetries and radial velocity shifts due to photospheric convective motions. The four components are made up of absorption line profiles calculated for granules, magnetic intergranular lanes, non-magnetic intergranular lanes, and magnetic bright points at disc centre. These components are constructed by averaging Fe I $6302~\mathrm{\AA}$ magnetically sensitive absorption line profiles output from detailed radiative transport calculations of the solar photosphere. Each of the four categories adopted are based on magnetic field and continuum intensity limits determined from examining three-dimensional magnetohydrodynamic simulations with an average magnetic flux of $200~\mathrm{G}$. Using these four component line profiles we accurately reconstruct granulation profiles, produced from modelling 12 x 12 Mm$^2$ areas on the solar surface, to within $\sim \pm$ 20 cm s$^{-1}$ on  a $\sim$ 100 m s$^{-1}$ granulation signal. We have also successfully reconstructed granulation profiles from a $50~\mathrm{G}$ simulation using the parameterised line profiles from the $200~\mathrm{G}$ average magnetic field simulation. This test demonstrates applicability of the characterisation to a range of magnetic stellar activity levels. 
\end{abstract}


\section{Introduction}
\label{sec:intro}
One of the consequences of plasma motions in the outer layers of low-mass stars with convective envelopes is radial velocity (RV) shifts due to variable stellar line profile asymmetries, known as astrophysical noise (or stellar `jitter'). This can pose a major problem for planet hunters as RV follow-up is mandatory for most planet confirmation and characterisation. As these net RV shifts, produced by photospheric convective motions, are on the m s$^{-1}$ level \citep{schrijver00}, stellar granulation will severely impact the confirmation of low-mass planets around solar-like stars. Additionally, the promise of cm s$^{-1}$ RV precision from spectrographs like Espresso and Codex makes it ever more pertinent that we understand granulation both in terms of the physical processes involved as well as an important source of astrophysical noise. 

\citet{meunier10a} conclude that the limiting factor for detecting Earth-like planets in the habitable zone is, in fact, the convective contribution to the RVs and not the larger amplitude contribution of spots and plages. \citet{meunier10a} go on to affirm that the amplitude of convective noise in solar-like stars, must  decrease by one order of magnitude or more to make possible the detection of low-mass planets. 

To date, there have been only a handful of attempts to decrease the effects of granulation for the benefit of exoplanet detection and confirmation. \citet{dumusque11a} propose adjustments to the number of measurements and exposure time in order to average out some of the effects of granulation. Although this method has had success (see \citealt{pepe2011}), it is both time and cost intensive, and does not provide information on the nature of granulation. \citet{aigrain12} take a different approach and use photometry to estimate RV variations due to stellar activity, including a parameter to estimate the perturbation due to convection. They successfully demonstrate that they can reproduce variations well below the m s$^{-1}$ level. However, as low-mass planets around solar-like stars require cm s$^{-1}$ precision, there is still an urgent need to further characterise and remove the RV variations of surface convection. 

Although there have been few attempts to remove or reduce the effects of granulation, there have been numerous observations and theories devoted to understanding the nature of granulation (for a comprehensive review see \citet{nordlund09}, \citet{stein12}, and references therein). We were particularly inspired by the four component model of \citet{dravins90}, used to reconstruct the observed asymmetries in stellar line profiles due to granulation. Similar to \citet{dravins90}, our aim is to measure and understand the RV signal produced by convective motions integrated across a stellar disc. However, our main goal is to simulate numerous granulation observations and ultimately develop a statistically significant noise removal method.

In this paper, the first in a series, we outline a technique to characterise granulation as an astrophysical noise source. The backbone of this characterisation is a state-of-the-art three-dimensional magnetohydrodynamic (MHD) solar simulation, coupled with detailed wavelength-dependent radiative transfer. Due to the time-intensive nature of detailed radiative diagnostics, producing enough realistic granulation patterns to cover an entire stellar disc with this method is not feasible. As such, we use the output from a short time-series of simulations to parameterise the granulation signal.

In Section~\ref{sec:comp}, we outline our procedure to separate absorption line profiles from solar simulations into four categories based on continuum intensity and magnetic field strength to categorise the different physical components present in photospheric magneto-convection. In Section~\ref{sec:recon}, we use these components to reconstruct the absorption line profiles from the simulated models. In Section~\ref{sec:50G} we show that this parameterisation can be successfully applied to other magnetic field strengths. We discuss our procedure to generate new realistic granulation line profiles in Section~\ref{sec:newgran} and present the success of our parameterisation as well as its current limitations in Section~\ref{sec:disc}. 

\begin{figure*}[ht]
\begin{center}
\includegraphics[width=\textwidth]{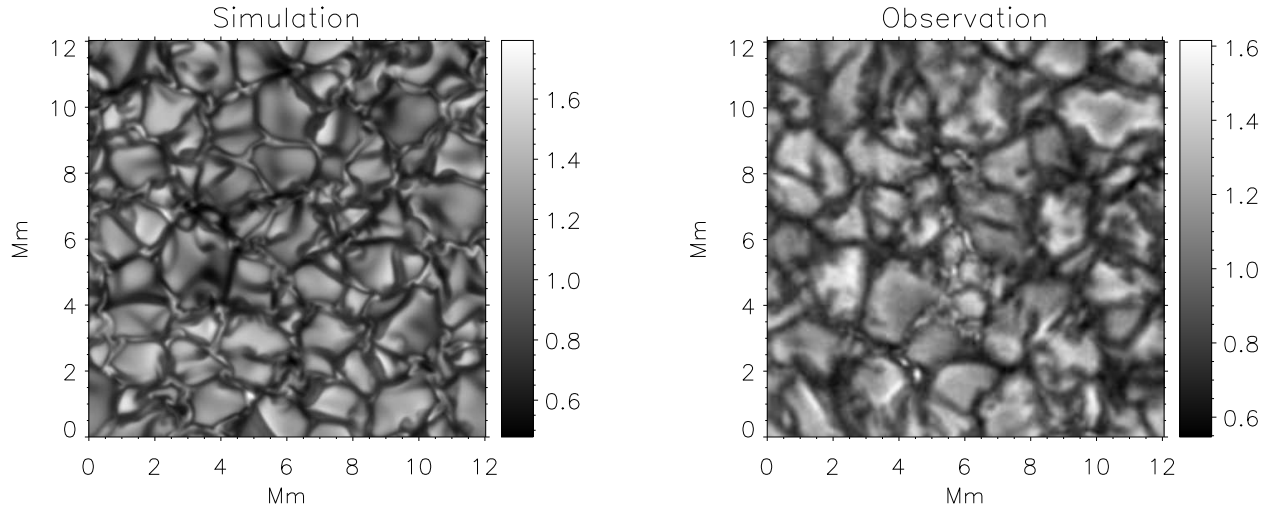}
\caption{An example of our simulation of solar granulation (left) in the white-light continuum (417 nm) compared to observations (right). The observations were taken with the Rapid Oscillation in the Solar Atmosphere (ROSA) imaging system on the Dunn Solar Telescope. The contrast in the observations is reduced due to seeing effects.} 
\label{fig:mhd}
\end{center}
\end{figure*}

\section{Characterisation of Convection Components}
\label{sec:comp}
The radiative MHD simulations of solar magneto-convection reproduce the properties of granulation observed at the solar surface in detail and for a wide range of magnetic activity levels (for a comparison of simulation and observation see Figure~\ref{fig:mhd} and for further comparisons with observables see \citealt{shelyag07}). However, there is no established to method to remove the granulation signal from observational data. We use such simulations, examining the Sun as a star in this paper, to develop a parameterised model of granulation that will eventually be used to remove convective noise from observations and can later be applied to other spectral types. Applicability to other spectral types is essential to developing a robust noise removal technique for exoplanet surveys due to the variety of planet-hosting stars. The simulations also provide many advantages towards developing such a technique including controllability of magnetic flux, seeing/resolution, and rotation. 

We simulate solar surface granulation using the MURaM code \citep{vogler05}. The code solves large-eddy radiative 3D MHD equations on a Cartesian grid, and employs a fourth-order central difference scheme to calculate spatial derivatives. The numerical domain has a physical size of 12 x 12 Mm$^{2}$ in the horizontal direction and 1.4 Mm in the vertical direction and is resolved by 480 x 480 x 100 grid cells (see \citealt{shelyag11} for details). 

We first introduce a uniform vertical $200~\mathrm{G}$ magnetic field and produce a sequence of 190 snapshots of solar magnetic granulation, with a $\sim$ 30 second cadence. The sequence covers approximately 80 minutes of physical time, corresponding to  $\sim$ 10 - 20 granular lifetimes. This does not include the first 20 minutes of the simulation, which is discarded, in order to allow the large amplitude oscillation, induced from the introduction of magnetic field, to dampen and the model to stabilise \citep{vogler05, vogler07}. During this first part of the simulation, the magnetic field is advected into the intergranular lanes by the convective plasma motions, and the maximum field strength rises from its initial value of $200~\mathrm{G}$ to a few kilogauss in the intergranular lanes. The $200~\mathrm{G}$ field is sufficiently strong to ensure each physical aspect of convection found in the quiet Sun is present in the simulations.

We also produce simulations in a similar manner for a uniform vertical $50~\mathrm{G}$ field. The cadence again is $\sim$ 30 seconds, and this simulation covers approximately 45 minutes of physical time. A similar oscillation in the $50~\mathrm{G}$ model is present throughout the simulation since the lower magnetic field does not provide as much damping compared to the $200~\mathrm{G}$ model. This weak magnetic field simulation provides a valuable test case for our parameterised model. An accurate reconstruction of the $50~\mathrm{G}$ model from the $200~\mathrm{G}$ components demonstrates the robustness of our parameterisation across magnetic field strengths.

The calculated models contain basic solar photospheric plasma parameters such as density, pressure, temperature, magnetic field strengths, and velocity. We perform the radiative transport diagnostics with these parameters, using a modified version of the STOPRO code (as described in \citealt{shelyag07}), for the simulated photosphere with the Fe I $6302~\mathrm{\AA}$ magnetically sensitive absorption line in local thermodynamic equilibrium (LTE) approximation. This line is widely used in both solar observations and simulations as a diagnostic tool for magnetic field and temperature \citep{shelyag07}. The full Stokes vector is calculated for each pixel in the snapshots throughout the simulation; for this study we use only the $I$ component of Stokes vector. The wavelength range used for calculation is $\pm~0.3~\mathrm{\AA}$ and is resolved by 400 wavelength points. 

\subsection{Four Component Model}
\label{sub:4comp}
The premise for our parameterisation of granulation line profiles is that they are a summation of absorption line profiles corresponding to each physical component of solar surface granulation. As such, we began our parameterisation by examining the different physical components of photospheric magneto-convection in the $200~\mathrm{G}$ simulations. Granules are non-magnetic, hot, bright bubbles of rising gas. The plasma in granules cools, darkens and falls back down into the intergranular lanes. The magnetic bright points (MBPs)  present within the dark intergranular lanes are due to the enhanced continuum intensity and decreased radiation absorption in  evacuated, highly magnetic flux tubes. Thus, a four component parameterisation, which divides the granulation pattern into granules, magnetic intergranular lanes, non-magnetic intergranular lanes and MBPs is a natural step to move forward. 

The simulation provides information on magnetic field and continuum intensity (an example of which is shown in the left panel of Figure~\ref{fig:mhd}), alongside the Stokes profiles. Since the four components differ inherently in brightness and magnetic field, it is logical to impose cutoff limits based on continuum intensity and magnetic field (measured at the continuum formation height) to separate each component within the simulation.  As a result, for each snapshot we average together the Stokes $I$ profiles that correspond to each category to create our component profiles. Thus, for each of the 190 snapshots in the time sequence there are four line profiles, each representing a component within the granulation pattern. 

To determine the optimal magnetic field and continuum intensity limits for the four components we explored the parameter space from 100 - $1200~\mathrm{G}$ and from 0.8 to 1.2 times the average continuum intensity. The final selection of limits, 0.9 average continuum intensity and $1000~\mathrm{G}$, was based on how well the four profile components could reconstruct the original simulated line profiles; the procedure of reconstruction is discussed in Section~\ref{sec:recon}. The four profiles used to characterise the different components of photospheric magneto-convection are shown in Figure~\ref{fig:4compall}. The breadth about each component's respective means is representative of the precision obtained. The characteristic bump seen in the MBP and magnetic intergranular lane profiles is due to the Zeeman splitting from the high magnetic fields present in these components. 

\subsection{Removing Oscillations}
\label{sub:osc}
Throughout the time-sequence, all profiles experience a RV shift. An examination of Fourier power in the pressure domain of the simulations, reveals two peaks at around 2 and 7 mHz. We attribute these variations to the solar p-modes that contribute to the well-known `5 minute' oscillations. In order to disentangle the granulation signal, this oscillation needs to be removed from the profiles. Removing the oscillation allows the convective motions to be viewed in detail and prevents time-average component profiles from being skewed by oscillation-induced broadening, thus allowing a more accurate granulation parameterisation.

\begin{figure}[h]
\includegraphics[width=8.5cm]{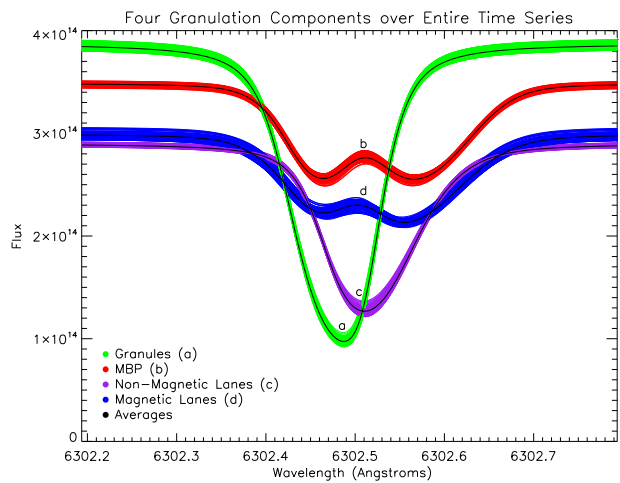}
\caption{Line profiles from the entire time series for the different physical components of granulation: granules (green), MBPs (red), magnetic (blue) and non-magnetic (purple) intergranular lanes. The time-average line profiles are shown in black. All components have been shifted by their bisector mean wavelength to the bisector mean wavelength of their time-average profile to remove the oscillation signal. The breadth about each component's respective means is representative of the precision obtained.} 
\label{fig:4compall}
\end{figure}

In order to remove the effects of the oscillation, we shifted each component profile to fixed reference points. The reference points were determined by first creating initial time-average component profiles. The mean values of the line bisector wavelengths from these initial time-average profiles served as the reference points because they had no knowledge of the oscillation. We then shifted the components by their bisector mean wavelengths to preserve shape variation due to the granulation, while simultaneously removing the oscillation effects. Once the profiles were shifted to remove the oscillation, we then used a final time-average for each component to be representative of each granulation feature. As seen in Figure~\ref{fig:4compall}, there is little variation within each component category throughout the time-sequence, providing confidence that the oscillation signal does not significantly affect component profile shape and that it has been successfully removed.

 \begin{figure*}[ht]
 \begin{center}
\includegraphics[height=10cm]{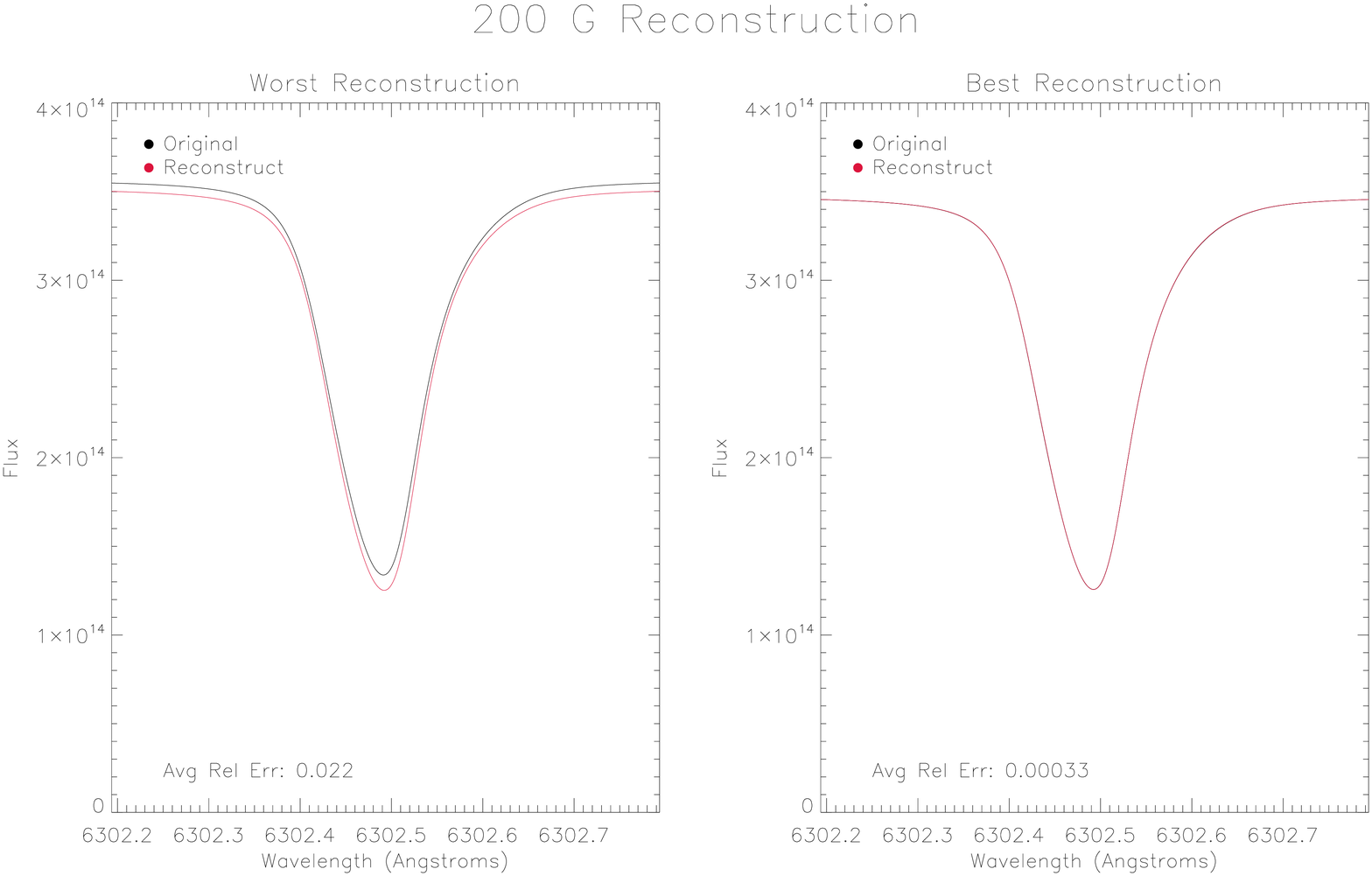} 
\caption{The reconstructed line profiles (red) are compared to the original line profiles (black) from the $200~\mathrm{G}$ simulations. Left: worst reconstruction achieved; right: best reconstruction achieved.} 
\label{fig:reconstruct}
\end{center}
\end{figure*}

\section{Parameterised Reconstruction} 
 \label{sec:recon}
We examined the accuracy of the four time-average line profiles, representing the different components of photospheric magneto-convection, by reconstructing the line profiles produced by the simulations. For each granulation snapshot, we recorded the proportion of the area occupied by each component to the total area (filling factor) based on the number of Stokes $I$ profiles in a given category to total number in the snapshot; mean values for these within $1\sigma$ are shown in Table~\ref{tab:fill}. To reconstruct the line profile for an entire snapshot, we multiplied each component profile by its corresponding filling factor and then added all four component profiles together. Examples of the reconstructions are shown in Figure~\ref{fig:reconstruct}.  
  
 \begin{deluxetable}{cc}
\tablecolumns{2}
\tablewidth{0pt}
\tabletypesize{\scriptsize}
\tablecaption{Mean filling factor for each component with $\pm 1 \sigma$}
\tablehead{\colhead{Component} & \colhead{Mean Filling Factor $\pm 1 \sigma$} }
    Granule & 0.559 $\pm$ 0.016  \\ 
    Non-Magnetic Intergranular Lane & 0.348 $\pm$ 0.019 \\ 
    MBP & 0.054 $\pm$ 0.004 \\ 
    Magnetic Intergranular Lane & 0.039 $\pm$ 0.006 \\ 
\label{tab:fill}
\end{deluxetable}
 
We calculated the average relative error, $\langle\Delta\mathrm{I}/\langle\mathrm{I}\rangle\rangle$ (where I is the continuum intensity), between the reconstructed profiles and the original profiles produced by the simulations. Even at the very worst, the reconstructed and original profiles still agree to at least $\sim$ 2\%, as seen in Figure~\ref{fig:reconstruct}. On average the reconstructed and original profiles are in agreement to $\sim$ 0.6\% or better.

 \begin{figure}[hb]
\includegraphics[width=8.5cm, height=0.4\textwidth]{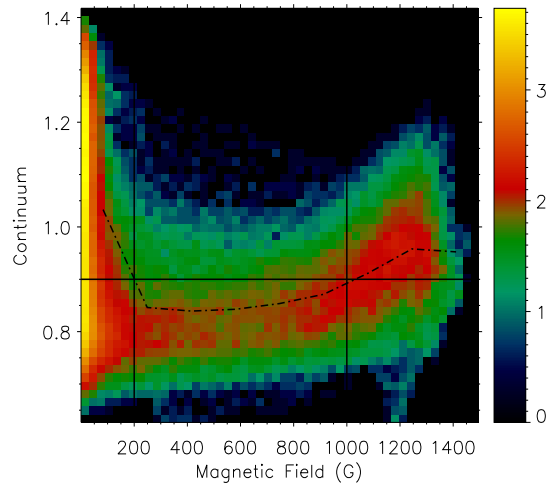}
\caption{The probability density distribution of the modulus of the vertical component of the magnetic field approximately at the continuum formation level versus continuum intensity for an arbitrarily chosen snapshot from the $200~\mathrm{G}$ simulation. The average dependence of the continuum intensity on the modulus of the vertical magnetic field strength is shown by the black dot-dashed curve. The cutoff limits for each of the physical granulation components are shown in black ($\sim$ 0.9 continuum intensity, $1000~\mathrm{G}$ for both four and six component models and, additionally, $200~\mathrm{G}$ for the six component model).}
\label{fig:bfieldcont}
\end{figure}

\subsection{Impact of Additional Components}
\label{sub:6comp}
 \begin{figure}[ht]
\includegraphics[width=8.5cm]{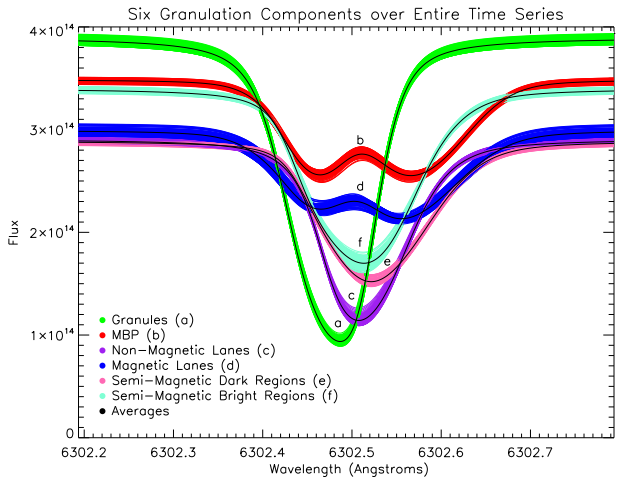}
\caption{Line profiles from the entire time series for the different physical components of granulation, as in Figure~\ref{fig:4compall}, with two additional components: granules (in green), MBPs (in red), magnetic (in blue) and non-magnetic (in purple) intergranular lanes, semi-magnetic bright (in turquoise) and semi-magnetic dark (in pink) regions. As before, oscillation effects have been removed and the breadth about each component's respective means (shown in black) is representative of the precision obtained.} 
\label{fig:6comp}
\end{figure}

We have also investigated the impact of assigning additional categories to our parameterisation scheme. Upon examining the probability density distribution of the modulus of the vertical component of magnetic field (measured roughly at the continuum formation layer in the simulation box) versus the continuum intensity, shown in Figure~\ref{fig:bfieldcont}, we realised a potential for an additional two components. There are two higher density clusters, one at high and one at low magnetic fields with a range of continuum intensities. These clusters indicate the presence of magnetic and non-magnetic features that are both bright and dark. In our four component model we separate these regions based on a magnetic field limit of $1000~\mathrm{G}$ and a continuum intensity limit of 0.9. However, it is clear from Figure~\ref{fig:bfieldcont} that many profiles with intermediate magnetic fields were being included in the non-magnetic components (granules and non-magnetic intergranular lanes). We chose to test a model where there was an additional magnetic field limit at $200~\mathrm{G}$ to separate the non-magnetic features from those with intermediate magnetic field. This additional limit then introduced two new components that we term semi-magnetic bright and semi-magnetic dark regions, shown in Figure~\ref{fig:6comp}, with magnetic field between $200~\mathrm{G}$ and $1000~\mathrm{G}$. We found that the six component model did not produce significantly smaller average relative errors when compared to the original line profiles from the simulations. In fact, at times the four component model even faired better, as seen in the top panel of Figure~\ref{fig:err_rv}. Even at the most extreme, the average relative errors between four and six component model differed by only $\sim$ 0.0005 ($<$ 1\%). For simplicity we proceed with the four component model. 

\subsection{Impact of Errors}
\label{sub:reconRV}
Ultimately we want to measure the RV signal produced by the convective motions integrated across a stellar disc. In order to do so, we must be confident in our ability to accurately reproduce the RVs from these convective motions using our four component parameterisation. We need to ensure that any resultant RV variations are not due to errors introduced in the reconstruction process. To do so we began by measuring the RVs of the reconstructed profiles. 

\begin{figure}[hb]
\includegraphics[width=8.5cm,height=11cm]{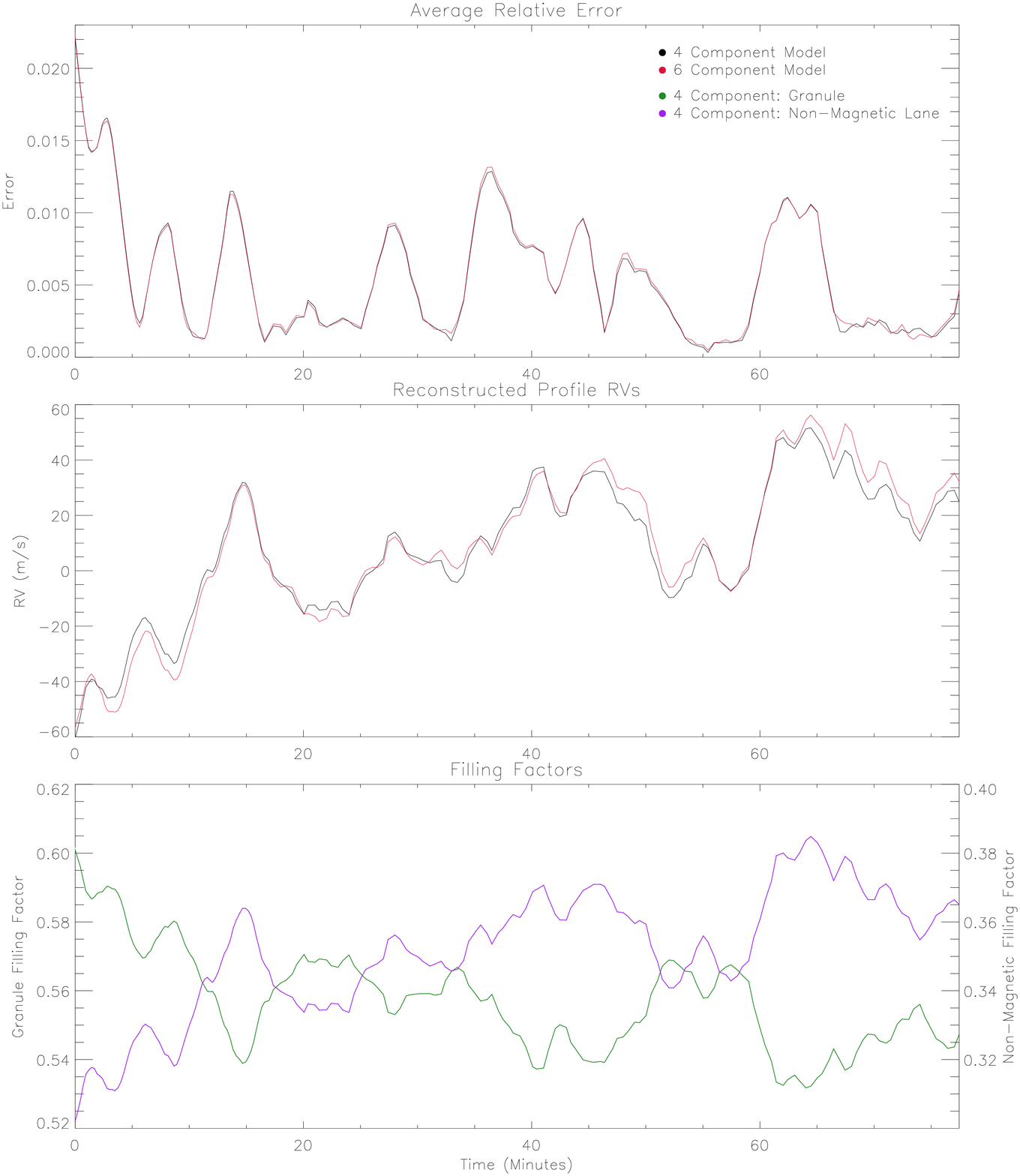}
\caption{Top panel: the average relative error between the original line profiles from the simulations and the reconstructed line profiles over time. Middle panel: the RVs of the reconstructed line profiles over time. Four component model is shown in black and six component model is shown in red. Bottom panel: filling factor vs time for the two largest components, granules (in green) and non-magnetic intergranular lanes (in purple) in four-component model. Any correlation between RV and average relative error is likely due to changes in filling factors that simultaneously impact the RV and line shape of a profile.} 
\label{fig:err_rv}
\end{figure}

We measured the RVs from reconstructed profiles by cross correlating a single, randomly selected profile with the entire time-series. At times, there is a possible correlation between average relative error (top panel, Figure~\ref{fig:err_rv}) and RV (middle panel, Figure~\ref{fig:err_rv}). We believe this is due to the relationship between filling factor and RV. For example, if the granule profile has a larger error, compared to the profiles from the other components, then a reconstruction with a higher granule filling factor would have a more negative RV and also a larger error. We see this in Figure~\ref{fig:err_rv}, where changes in component filling factor (bottom panel, shown in green for granules and in purple for non-magnetic intergranular lanes) correlate with RV (middle panel) and error (top panel). However, such a correlation could also indicate that the measured RVs may originate from errors within the reconstruction process rather than the granulation signal.

\begin{figure*}
\begin{center}
\includegraphics[width=\textwidth, height=11cm]{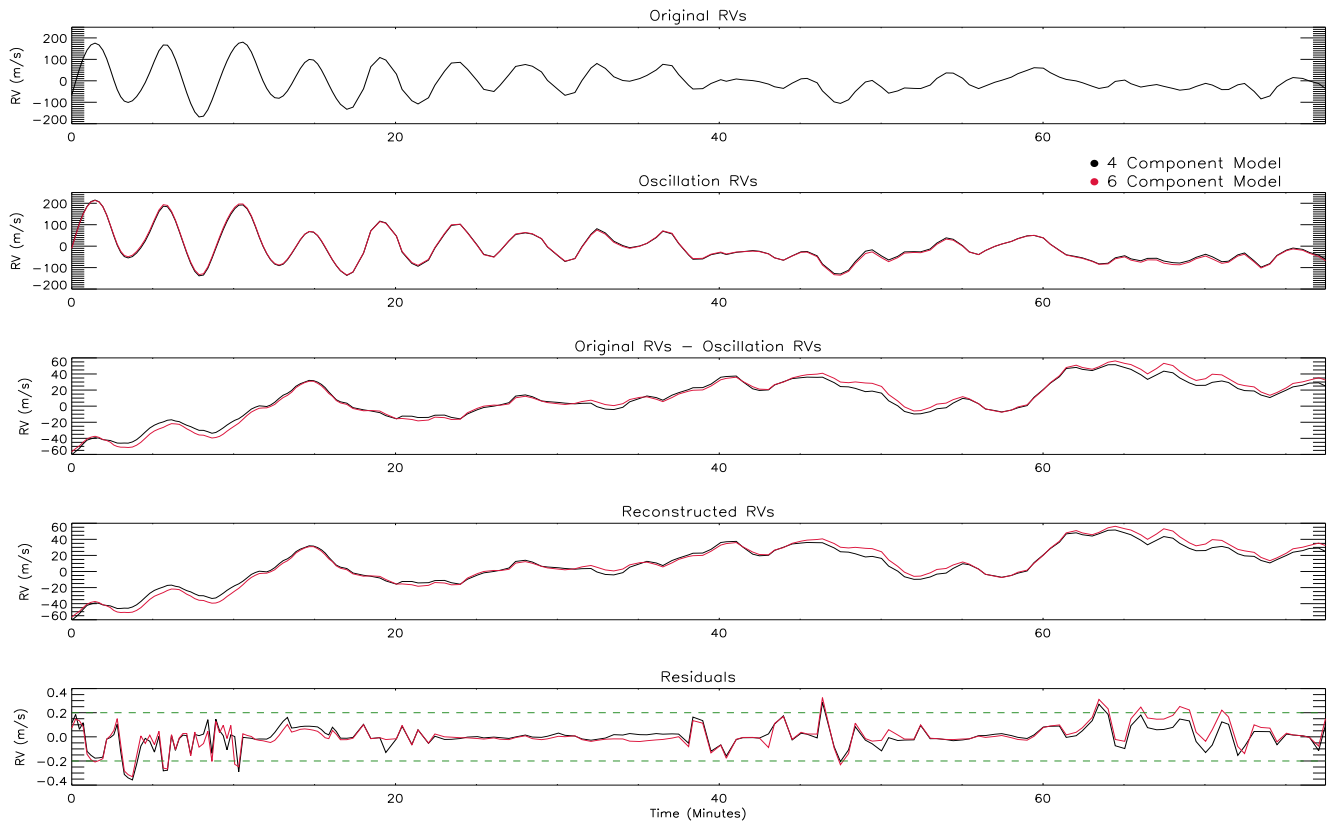}
\caption{First panel: RVs of the original profiles over the time series, the oscillation is clearly evident in the first half of the series. Second panel: RVs of the approximate oscillation signal. Black and red curves represent use of the four and six component reconstructions, respectively. Third panel: original profile RVs after approximate oscillation signal has been subtracted. Fourth panel: RVs of the reconstructed profiles. Fifth panel: difference between original and reconstructed RVs, after removal of oscillation. As shown in the fifth panel, our reconstruction is good to $\sim \pm$ 20 cm s $^{-1}$ on a $\sim$100 m s$^{-1}$ granulation signal (green dashed lines are plotted only to guide the eye).} 
\label{fig:orgoscrecon}
\end{center}
\end{figure*}

To confirm that the measured RVs for the reconstructed profiles do originate from the granulation signal, and not from reconstruction errors, we compared the reconstructed RVs to those of the original line profiles. To measure the RVs of the original profiles we cross-correlated them with a randomly selected reconstructed profile. We used a reconstructed profile, arbitrarily chosen, as a reference because it has no knowledge of the oscillation (first panel, Figure~\ref{fig:orgoscrecon}).  As mentioned in Section~\ref{sub:osc}, we removed the oscillation for our parameterisation and as a result we must also remove it from the original profiles to directly compare with the reconstructed profiles. To do so, we began by estimating the oscillation signal. If the reconstruction profiles match perfectly the simulated granulation then a cross-correlation of the original profiles with their respective reconstructed profiles would provide the oscillation signal alone. While the reconstruction process is not perfect, we assume it is good enough to give a good approximation of the oscillation signal (second panel, Figure~\ref{fig:orgoscrecon}). This approximate oscillation signal is then subtracted from the original profile RVs, and therefore any remaining RV signal should be due to the convective motions and any reconstruction errors (third panel, Figure~\ref{fig:orgoscrecon}. We note here that if the oscillation signal were to significantly alter the component profile shape, then remnant signatures would be present after this stage and in the reconstructions, which we do not find.). We then compared this remaining RV signal to the reconstructed RVs (fourth panel, Figure~\ref{fig:orgoscrecon}), and calculated the difference between them to produce the residuals (fifth panel, Figure~\ref{fig:orgoscrecon}). The residuals are approximately $\pm$ 20 cm s$^{-1}$ for more than 95\% of the time-sequence. As such, we conclude that our categorisation is good to $\pm$ 20 cm s$^{-1}$ on a $\sim$ 100 m s$^{-1}$ granulation RV variation calculated over a 12 x 12 Mm$^2$ area. We therefore conclude that any errors in the parameterisation do not significantly impact our ability to generate line profiles that represent realistic granulation patterns. 

 \begin{figure*}
 \begin{center}
\includegraphics[height=10cm]{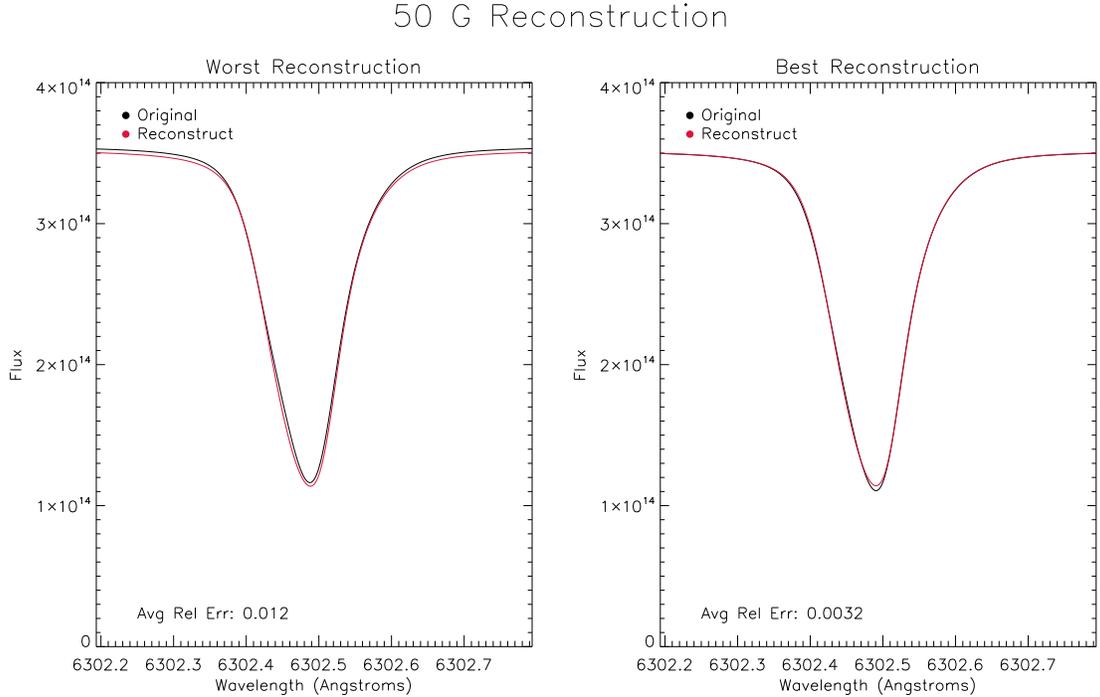}
\caption{The four component line profiles from $200~\mathrm{G}$ magnetic field simulation are used in conjunction with filling factors from $50~\mathrm{G}$ simulation to reconstruct granulation line profiles in a $50~\mathrm{G}$ field. The reconstructed line profiles (in red) are compared to the original line profiles (in black). Left: worst reconstruction achieved; right: best reconstruction achieved.} 
\label{fig:50g}
\end{center}
\end{figure*}
 
\section{$50~\mathrm{G}$ Reconstruction}
\label{sec:50G}
Our four component model parameterises granulation line profiles from the $200~\mathrm{G}$ simulations, where the full spectrum of photospheric magneto-convection features is present. In accordance with our premise (Section~\ref{sub:4comp}), such a parameterisation should be capable of accurately reproducing granulation line profiles throughout the quiet Sun, where average surface magnetic fields vary from near 0 up to $\sim$ $200~\mathrm{G}$.  We confirm this by using our four component $200~\mathrm{G}$ parameterisation to reconstruct granulation profiles produced from the simulation with a $50~\mathrm{G}$ average field. 

The $50~\mathrm{G}$ simulation was parameterised in the same manner as the $200~\mathrm{G}$ (see Section~\ref{sub:4comp}). From the simulation and parameterisation, we found the filling factors of each of the four granulation components in a $50~\mathrm{G}$ model. As such, we use these filling factors in conjunction with the four component profiles from the $200~\mathrm{G}$ parameterisation to reconstruct the profiles from the $50~\mathrm{G}$ simulations. As shown in Figure~\ref{fig:50g}, we have achieved excellent accuracy in our reconstruction, ranging from at worst $\sim$ 1\% to at best $\sim$ 0.3\%. From these results, we are confident that our parameterisation of Stokes $I$ profiles is robust for the various magnetic field strengths found in the quiet
 Sun.

\section{Generating New Granulation Line Profiles}
\label{sec:newgran}
\begin{figure}[h]
\includegraphics[width=8.5cm]{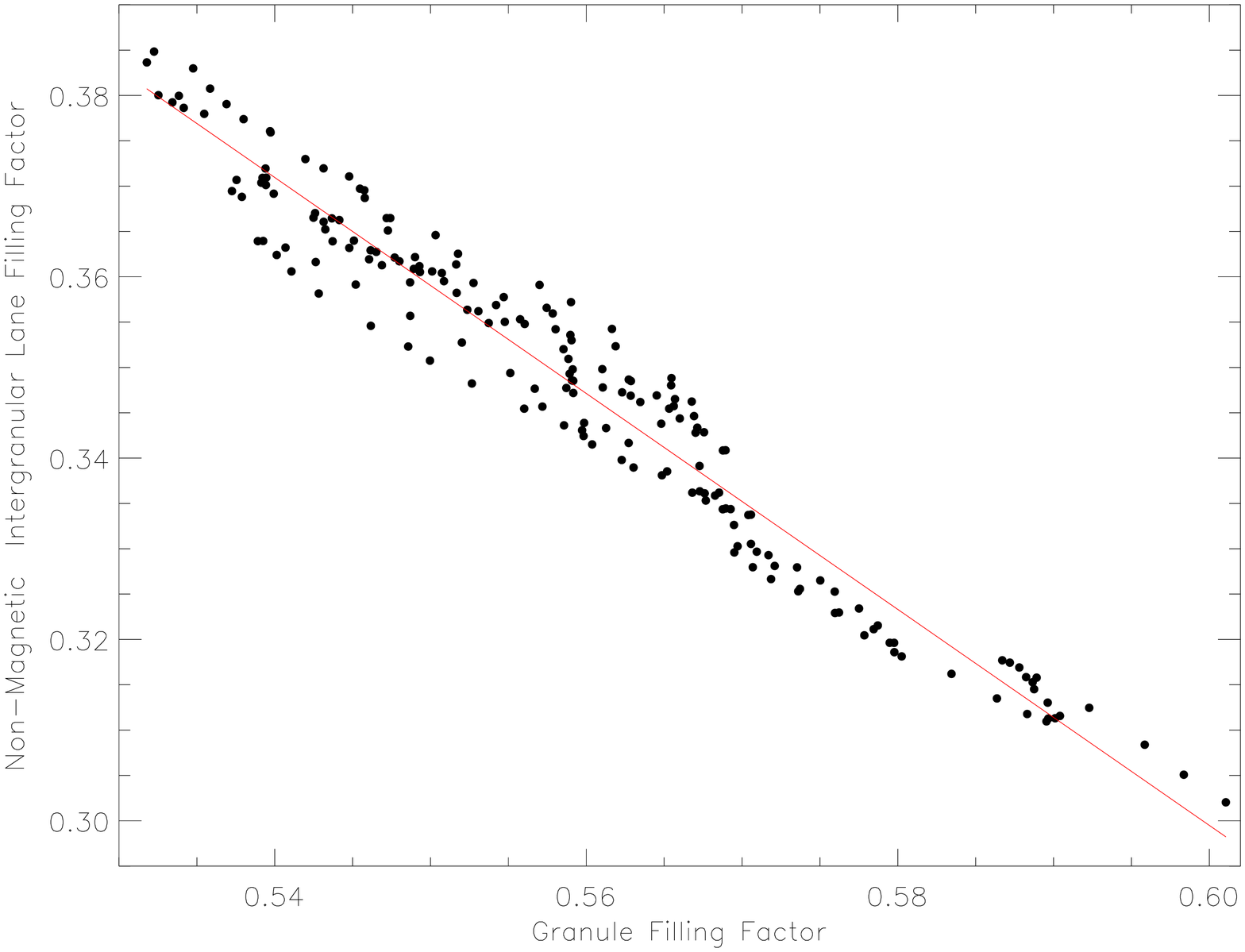}
\caption{There is a strong anti-correlation between granule filling factor and the non-magnetic intergranular lane filling factor. The solid red curve is a linear fit. This fit is used in conjunction with a randomly selected granule filling factor to determine the non-magnetic intergranular lane filling factor used to generate new stellar line profiles representative of realistic granulation patterns.} 
\label{fig:fillingfactor}
\end{figure}

To create new stellar line profiles that represent realistic granulation patterns over a 12 x 12 Mm$^{2}$ patch, we generate a probability distribution for the granules based on filling factors throughout the time-sequence. This allows us to randomly and realistically select the granule filling factor for a given profile. As seen in Figure~\ref{fig:fillingfactor}, there is a strong linear anti-correlation (with slope -1.19) between granule and non-magnetic intergranular lane filling factor (with a Pearson's correlation coefficient of -0.98). As a result, the granule filling factor is used in conjunction with this anti-correlation to determine the filling factor for the non-magnetic intergranular lanes. Together, granule and non-magnetic intergranular lanes make up approximately 90\% of the total granulation profile (Table~\ref{tab:fill}). Since there are no other strong correlations amongst filling factors, a probability distribution is again generated, this time for the magnetic intergranular lane filling factor. As the four component filling factors must total to one, the MBP filling factor is then determined by the difference between the other three filling factors and one. Once all four filling factors are determined we use these in conjunction with our four component profiles determined above to create new realistic granulation line profiles.

\section{CONCLUDING REMARKS}
\label{sec:disc}
Simulating realistic solar granulation can take up to months for a sequence of snapshots that last little more than an hour. Computing the Stokes profiles from these simulations for just one snapshot can take hours on its own right. Currently, it is not feasible to tile a stellar disc with each 12 x 12 Mm$^{2}$ tile having an independent granulation pattern produced by simulations. However, using our parameterised model described here we can produce accurate granulation line profiles for magnetic field scenarios between 50 - $200~\mathrm{G}$. In addition, this method reproduces radial velocities which are accurate to $\sim \pm$ 20 cm s$^{-1}$ on a $\sim$100 m s$^{-1}$ granulation signal.

Throughout this paper we outlined the methodology for our parameterisation of granulation at disc centre. Naturally, the granulation component characterisation may change as we apply our parameterisation to various disc positions, however, our methodology will remain the same. Applying our parameterised method to off-centre regions of the stellar disc is more difficult. Photospheric magneto-convection results in a corrugated surface. An apt analogy is to think of the granules as `hills' and the intergranular lanes as `valleys'  \citep{dravins08}. Beyond zero degrees, we begin to see different aspects of these `hills' and `valleys.' As we incline the simulations the granular walls become visible, while the intergranular lanes become ever more hidden, and MBPs disappear from view toward the limb. Consequently, there are changes in profile shape for granulation patterns at various inclinations. Furthermore, higher inclinations lead to a physically longer line-of-sight and require finer resolution (in terms of grid cell number) along the inclined line-of-sight due to appearance of strong under-resolved gradients in plasma parameters. Thus, in order to maintain the same precision, an even greater amount of computational time is required. In a future paper, we will extend our parameterisation across the stellar disc using the methodology outlined here. 

Once we have a model that applies to various stellar disc positions, we will tile  a stellar disc with line profiles that represent random, independent granulation patterns and integrate the wavelength-dependent intensity over the disc. These `simulated star' observations will allow us to search for a way to characterise the granulation signal as a whole by examining relationships between line asymmetries, bisector changes and RV. As this method stems from the radiative 3D MHD simulations, we plan to expand these simulations to other spectral types thereby allowing our parameterised method to be tested beyond the solar spectrum. In future work, we will also explore such a parameterisation and search for noise removal with other stellar absorption lines--we begin here with one line to validate our method. There are a number of potential routes for achieving RV detection of Earth-like planets. Currently, RV measurements of exoplanets are done via cross-correlation with a spectral type template mask, requiring several thousand lines. However, it may be that for Earth-like confirmation we may have to target specific, well-behaved unblended lines, in which case we may only need a handle of lines modelled. Additionally, if we do find correlations between granular noise and observables then we may be able to search for said correlations empirically for a variety of lines, which may limit the need for such detailed modelling of other stellar lines. 

With our current parameterised four component model, we are able to accurately match the synthesised Stokes $I$ profiles obtained from radiative diagnostics of 3D MHD radiative simulations of the solar photosphere and mimic observations of solar granulation, while maintaining the opportunity to apply our model to other spectral types. As we move ever closer towards the confirmation of Earth-like worlds, granulation and other low-level noise sources such as meridional flows or variable gravitational redshift \citep{beckers07, makarov10, cegla12} become an imminent limitation. While, our primary goal is to remove/reduce granulation effects in the exoplanet searches, a characterisation of granulation is multidisciplinary and may be beneficial for solar physics and stellar astrophysics. Furthermore, as we expand our approach to other spectral types we will not only be able to assist exoplanet surveys around a larger variety of host stars, but also understand the convection process more thoroughly across the HR diagram.  

\section*{\sc Acknowledgments}
This research has made use of NASA's Astrophysics Data System Bibliographic Services. HMC acknowledges support from a Queen's University Belfast university scholarship. CAW would like to acknowledge support by STFC grant ST/I001123/1. Financial support from the Leverhulme Trust (grant RPG-249) is gratefully acknowledged. We thank the anonymous referee for their comments which have significantly helped improve the clarity of this paper. 

\bibliographystyle{apj}
\bibliography{abbrev,refs}

\end{document}